\documentstyle[11pt,epsf]{article}
\newfont{\bg}{cmr10 scaled\magstep4}
\newcommand{\bigzerol}{\smash{\hbox{\bg 0}}}
\newcommand{\bigzerou}{\smash{\lower1.7ex\hbox{\bg 0}}}
\newcommand{\dotmatrix}[1]{\multicolumn{#1}{c}{\dotfill} \\}
\newcommand{\teigi}{\stackrel{\rm def}{=}}
\newcommand{\sump}{\mathop{{\;\,{\sum}'}}}
\newcommand{\NN}{\nonumber}

\newcommand{\expect}[1]{\langle{#1}\rangle}

\newcommand{\BE}{\begin{equation}}
\newcommand{\EE}{\end{equation}}
\newcommand{\BEA}{\begin{eqnarray}}
\newcommand{\EEA}{\end{eqnarray}}
\newcommand{\BEAN}{\begin{eqnarray*}}
\newcommand{\EEAN}{\end{eqnarray*}}

\newcommand{\IE}{{\it i.e.}~}
\newcommand{\CHU}[1]{~{\rm{#1}}~}

\newcommand{\ANN}[3]{Ann. Phys. (NY) {\bf #1} {(#2)} {#3}}

\newcommand{\IJMP}[3]{Int. J. Mod. Phys. {\bf #1}{(#2)}{#3}}

\newcommand{\JP}[3]{J. Phys. {\bf #1} {(#2)} {#3}}
\newcommand{\JPSJ}[3]{J. Phys. Soc. Jpn. {\bf #1} {(#2)} {#3}}
\newcommand{\JSP}[3]{J. Stat. Phys. {\bf #1} {(#2)} {#3}}

\newcommand{\PR}[3]{Phys. Rev. {\bf #1} {(#2)} {#3}}
\newcommand{\PRL}[3]{Phys. Rev. Lett. {\bf #1} {(#2)} {#3}}

\newcommand{\LR}{\left(}
\newcommand{\RR}{\right)}

\begin{document}
\begin{center}
{\large{\bf Integrable Boundary Conditions in Asymmetric Diffusion Processes}}
\vskip 0.2cm
{\large {\bf{Akira Fujii}\footnote{e-mail:{\sf af@thp.uni-koeln.de}}}
\vskip 0.2 cm
{\small {\sl Universit{\"a}t zu K{\"o}ln, Institut f{\"u}r Theoretische Physik,
      50937 K{\"o}ln, Germany}}}
\vspace{0.5cm}
\end{center}
\begin{abstract}
We study the asymptotic diffusion processes with (generally nonlocal) 
open boundaries in one dimension which are exactly solvable 
by means of the recently developed {\it recursion formula}. 
We investigate the stationary states, which cannot 
be determined in an elementary way. We give the equation 
which includes an auxiliary parameter and determines 
possible boundary conditions 
for the model to be solved exactly. 
With the help of that equation, we analyze the density current
and concentration. We classify 
the phases according to them. 
\end{abstract}
%
\section{Introduction}\label{s:Introduction}
The study of the systems with reaction and diffusion 
in one dimension has been an attracting problem. 
or example, models including both reaction and diffusion 
have been considered in view of chemical processes like 
$A+B\rightarrow C+D$, where $A$,$B$,$C$ and $D$ represent 
some sorts of molecules. Models with only pure diffusion are not 
trivial, either. They have been studied in relation with interface growth, 
traffic flow and so on\cite{liggett}. 
Some experiments corresponding with 
the recent theoretical developments 
have been proceeded\cite{experiment}. Recently, a problem 
concerning protein synthesis has been considered in this 
scheme\cite{RNA}. 
To investigate reaction-diffusion models, a variety of methods 
have been used. Once taking a continuum limit, traditional technique of 
the field theory such as renormalization group method, mean-field or 
variational approximation are available\cite{doi,KPZ,RG}. 

In this paper, we consider the models on a one-dimensional discretized 
lattice. Some of the models on a lattice are shown to be integrable 
by means of the Jordan-Wigner transformation, Bethe ansatz\cite{gwa} 
and so on. 
%
%

The definitions and notations of reaction-diffusion 
models considered in this paper are as follows.   
We consider a one-dimensional lattice, where lattice sites are numbered 
1,2,$\cdots$,$L$(= the number of sites), and several sorts of particles 
living on the lattice. The total number of particles 
is not larger than that of lattice sites, \IE empty sites represented 
by $\phi$ are allowed. The stochastic process obeys 
the following continuous {\rm flow}~{\rm chart}. 
\begin{enumerate}
\item\label{en:choice} 
At an instant $t$, a pair of neighboring sites, 
say the $i$-th and $(i+1)$-th sites, is chosen randomly. 
We assume that the sorts of particles 
on the $i$-th and $(i+1)$-th sites are $A$ and $B$, respectively.
\item\label{en:reaction} 
The particles on $i$ and $i+1$ ($A$ and $B$) react into 
the same or other sorts of particles $C$ and $D$ respectively 
in an infinitesimal time interval $dt$ with a given probability 
$w^{i}_{C-A,D-B}(C,D)dt$. This process is represented by 
$A+B\rightarrow C+D$ and we must 
tell the difference of  
the order of particles (or sites), 
\IE $A+B$ is different from $B+A$ in general.
\item\label{en:timeevo}The time should be set $t\rightarrow t+dt$. 
\item A process \ref{en:choice}-\ref{en:timeevo} is repeated 
until $t$ amounts to the final time $t_{f}$ defined properly. 
We often put $t_{f}=+\infty$. 
\end{enumerate}
In this paper, we concentrate in models 
including only particles of one species, $A$,  
and the parameters $\mu$,$\nu$,$\alpha$ and $\beta$ in 
$w^{i}_{\alpha,\beta}(\mu,\nu)$ being $\bf{Z}_{2}$ variables. 
Because of the preservation of the probability, we have the relation 
\BE 
w_{0,0}^{i}(\mu,\nu)=\sump_{r,s}w_{r,s}^{i}(\mu-r,\nu-s),\label{eq:conpro}
\EE
hereafter $\displaystyle{\sump_{n_{1},n_{2},\cdots}}$ means 
that $n_{1}=n_{2}=\cdots=0$ in the summation is excluded. 
Let us introduce the probability distribution 
$P_{L}(\tau_{1},\cdots,\tau_{L};t)$ 
meaning the probability to find the $i$-th site to be occupied by $A$ 
(empty) if $\tau_{i}=1$ (0) at the time $t$. The time evolution equation 
of $P_{L}$ is a linear one denoted as 
\begin{equation}
{\partial\over\partial t}P_{L}(\tau_{1},\cdots,\tau_{L};t)=
({\hat H}P_{L})(\tau_{1},\cdots,\tau_{L};t),
\end{equation}
where ${\hat H}$ is a linear operator called {\it Hamiltonian} on the 
analogy of the Schr{\"o}dinger equation. Needless to say, $H$ can 
be written in terms of the local interactions characterized by 
$w$'s as we will see later. We consider the ground state, a solution 
of ${\hat H}P_{L}=0$, which the system will reach after a long time. 

Especially, we deal with problems of the stationary 
state of an open chain with pure asymmetric exclusion processes (ASEP). 
Although 
the setting is quite simple, to obtain the exact form of the 
stationary state is not easy. 
With some classes of models, useful recursion relations 
to determine physical quantities were indicated by Derrida, Domany and 
Mukamel (DDM)\cite{DDM}. They solved the recursion relation for a 
special case and predicted the phase diagram by means of the 
mean-field approximation. The solutions 
with general {\it local} boundary condition were obtained by 
Sch{\"u}tz and Domany (SD) without using the mean-field approximation
\cite{SD}.
 
In this paper, we will 
consider general boundary conditions including nonlocal boundary interactions 
so that a recursion 
relation similar to DDM's might be available, \IE integrable. 
We calculate some physical quantities, the partition function, density 
current and concentration (one-point function), exactly and classify 
the {\it phases} of the system. The plan of this paper is as follows. 
In the following \S\ref{s:boundary}, we have a brief sketch of 
the previous works by DDM and SD.  
An equation for the possible integrable boundary conditions is 
given in \S\ref{s:gbc}. We calculate the partition function, density 
current and concentration in \S\ref{s:pf} and \S\ref{s:phase}. 
An unknown example is analyzed in \S\ref{s:v_2}. Section \ref{s:conclusion} 
is left for the summary and conclusion.

\section{Asymmetric Exclusion Process}\label{s:boundary}
In this section, we consider pure-diffusion systems with boundaries, 
\IE open chains. Although it seems to be very simple, even determination 
of the ground state is not easy. 
An epoch-making work in this field 
was done by DDM\cite{DDM}. 
They made use of a 
recursion relation among the ground state with different lattice length, 
and obtained the solution for a special case ($\alpha=\beta=1$ in the 
later notation). The phase diagram classified by the strengths of the 
boundary injection and subtraction rates were conjectured by means of 
the mean-field approximation. SD proved DDM's conjecture to be true by 
obtaining the exact form of the solutions for the general boundary conditions. 
Another remarkable succeed is development of 
the so-called {\it matrix} {\it product} {\it method} introduced in 
ref.\cite{DEHP}. This method enables correlation functions 
to be calculated algebraically and exactly. Several extension 
can be found in refs.\cite{Sandow,ER,SS,sasamotowadati,niko}, 
for example. Recently, this method 
has been applied for the models including coagulation and decoagulation 
as well as pure diffusion\cite{Weizmann}. 

In this section, 
we concentrate in the simplest ASEP model on a lattice with 
length $L$ \cite{DDM} 
defined as follows. 
\begin{enumerate}
\item Left edge ($i=1$)\\
At the left edge, the injection and extraction of a particle are possible 
\[ \Phi\rightarrow A\qquad\CHU{with}\CHU{a}\CHU{rate}w_{1}(1)=\alpha\]
or
\[ A\rightarrow \Phi\qquad\CHU{with}w_{1}(0)=\gamma.\]
\item Bulk property ($i=1,\cdots,L$)\\
Only diffusion is allowed in the bulk like 
\[ A+\Phi\rightarrow\Phi +A\qquad\CHU{with}w_{11}(01)=p\]
or
\[ \Phi+A\rightarrow A+\Phi\qquad\CHU{with}w_{11}(10)=q.\]
\item Right edge ($i=L$)\\
At the right edge, the injection and extraction of 
a particle are possible 
as well as at the left edge.
\[ \Phi\rightarrow A\qquad\CHU{with}w'_{1}(1)=\delta\]
or
\[ A\rightarrow \Phi\qquad\CHU{with}w'_{1}(0)=\beta.\]
\end{enumerate}
We assume the conservation of the probability, for example 
\[
\Phi\rightarrow\Phi\qquad\CHU{with}1-\alpha
\]
at the left edge. 
\begin{figure}
  \epsfxsize = 6 cm   
  \centerline{\epsfbox{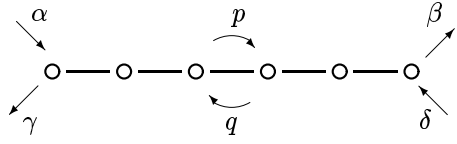}}
  \caption{Pure diffusion model with open boundaries.}
\label{f:hopping}
\end{figure}
The Hamiltonian of this model can be shown to be equivalent to that 
of $XXZ$ spin chain with integrable boundary terms\cite{ER,ADHR}.  
In this section, we consider the case with $q=\gamma=\delta=0$ 
for simplicity. 
Considering the rescaling of the time, we can put $p=1$ without any 
loss of the generality. 
We are interested in the stationary solution of the equation 
\BEA
0&=&
\sum_{\sigma_{1}=0,1}(h_{1})^{\sigma_{1}}_{\tau_{1}}
P_{L}(\sigma_{1},\tau_{2},\cdots,\tau_{L}) \NN\\
&&+\sum_{i=1}^{L-1}\sum_{\sigma_{i},\sigma_{i+1}=0,1}
h^{\sigma_{i}\sigma_{i+1}}_{\tau_{i},\tau_{i+1}}
P_{L}(\tau_{1},\cdots,\tau_{i-1},\sigma_{i},\sigma_{i+1},
\tau_{i+2},\cdots,\tau_{L})\NN\\
&&+\sum_{\sigma_{L}=0,1}(h')^{\sigma_{L}}_{\tau_{L}}
P_{L}(\tau_{1},\cdots,\tau_{L-1},\sigma_{L}), \label{eq:statma}
\EEA
where
\BEA
h_{1}&=&\LR\begin{array}{cc} \alpha&-\gamma\\ -\alpha&\gamma\end{array}\RR
=\LR\begin{array}{cc} \alpha&0\\ -\alpha&0\end{array}\RR, \label{eq:led}\\
h&=&\LR\begin{array}{cccc}
0&0&0&0 \\ 0&p&-q&0 \\ 0&-p&q&0 \\ 0&0&0&0 \end{array}\RR\NN\\
&=&\LR\begin{array}{cccc}
0&0&0&0 \\ 0&1&0&0 \\ 0&-1&0&0 \\ 0&0&0&0 \end{array}\RR,\label{eq:bpure}\\
h'&=&\LR\begin{array}{cc}\delta&-\beta\\ -\delta&\beta\end{array}\RR
=\LR\begin{array}{cc}0&-\beta\\ 0&\beta\end{array}\RR. \label{eq:red}
\EEA
The readers should remember that we do not have to restrict parameters 
less than one 
once we write the master equation in the above form by rescaling 
the time. We can consider that they are arbitrary positive numbers. 

In DDM's and the following work, it has been found that  
that the solution of eq.(\ref{eq:statma}) 
with general $p$,$q$,$\alpha$,$\beta$,$\gamma$ and $\delta$ 
satisfies surprising recursion relations for $L$ as follows. 
\BEA
\hspace{-1.0cm}&&\sum_{\sigma_{1}=0,1}(h_{1})^{\sigma_{1}}_{\tau_{1}}
P_{L}(\sigma_{1},\tau_{2},\cdots,\tau_{L})
=-x_{\tau_{1}}P_{L-1}(\tau_{2},\cdots,\tau_{L}), \label{eq:lrec}\\ 
\hspace{-1.0cm}&&\sum_{i=1}^{L-1}\sum_{\sigma_{i},\sigma_{i+1}=0,1}
h^{\sigma_{i}\sigma_{i+1}}_{\tau_{i},\tau_{i+1}}
P_{L}(\tau_{1},\cdots,\tau_{i-1},\sigma_{i},\sigma_{i+1},
\tau_{i+2},\cdots,\tau_{L})\NN\\ 
\hspace{-1.0cm}&&\quad=x_{\tau_{i}}
P_{L-1}(\tau_{1},\cdots,\tau_{i-1},\tau_{i+1},\cdots,\tau_{L})
-x_{\tau_{i+1}}
P_{L-1}(\tau_{1},\cdots,\tau_{i},\tau_{i+2},\cdots,\tau_{L}), 
\label{eq:brec}\\ 
\hspace{-1.0cm}&&\sum_{\sigma_{L}=0,1}(h')^{\sigma_{L}}_{\tau_{L}}
P_{L}(\tau_{1},\cdots,\tau_{L-1},\cdots,\sigma_{L})
=x_{\tau_{L}}P_{L-1}(\tau_{1},\cdots,\tau_{L-1}), \label{eq:rrec}
\EEA
where $x_{0}$ and $x_{1}$ are some numbers, which can be set as 
$-x_{0}=x_{1}=1$. It 
should be remarked that the normalization property of 
the probability, $\sum P_{L}(\sigma_{1},\cdots,\sigma_{L})=1$ 
is no more correct. 
We must renormalize the distribution 
function after obtaining its form\cite{DDM,SD}. 
We consider the normalization, \IE partition function, in \S\ref{s:pf} 
in detail. 

With the help of this recursion, we can calculate 
the correlation functions, for example the concentration, with arbitrary 
$L$ in principle. Contrary to the calculation of the concentration, those 
of general $n$-point function are difficult, in fact. Some technology 
to calculate the general correlation functions 
like the {\it matrix}~{\it product}~{\it method}\cite{DEHP} 
have been developed. 
%
\section{General Boundary Condition}\label{s:gbc}
As we have seen in the previous section, the recursion relation 
(\ref{eq:lrec})-(\ref{eq:rrec}) will play a crucial 
role to solve the model 
exactly. We consider this recursion relation in detail. 
In general, we consider a system denoted symbolically by 
\[
\bigcirc-\circ-\overbrace{\circ-\circ-\cdots-\circ-\circ}^{n}
, \quad n=0,1,2,\ldots. 
\]
Here we introduce an $s$-valued variable $X(=P_{1},\ldots,P_{s})$ living 
on $\bigcirc$ at the left edge and a usual 2-valued variable 
$\sigma(=0,1)$ on $\circ$.
Let us denote $X\in V_{s}$ and $\sigma\in V_{2}$ from now. The connection 
$-\circ-\circ-$ 
represents a pure asymmetric hopping, which we put $p=1$ and $q=0$ 
in this section. 
At the right edge ($-\circ$), 
a particle is extracted with $\beta=1$ (,$\delta=0$).

Denoting the distribution function for the ground state as 
$f_{n}(X,\sigma_{0},\cdots,\sigma_{n})$, 
the master equation takes the following form
\BEA\hspace{-0.8cm}
0&=&\sum_{Y\in V_{s}}\sum_{\rho_{0}}H^{Y\rho_{0}}_{X\sigma_{1}}
f_{n}(Y,\rho_{0},\sigma_{1},\ldots,\sigma_{n})\NN\\  
&&\! +\sum_{i=0}^{n-1}
h^{\rho_{i}\rho_{i+1}}_{\sigma_{i}\sigma_{i+1}}
f_{n}(X,\sigma_{0},\ldots,\sigma_{i-1},\rho_{i},\rho_{i+1},\sigma_{i+2},
\ldots,\sigma_{n})+(h')^{\rho_{n}}_{\sigma_{n}}
f_{n}(X,\sigma_{0},\ldots,\sigma_{n-1},\rho_{n})\NN\\ &&
\label{scho}
\EEA
with the {\it boundary matrix} $H^{Y\rho}_{X\sigma}$, which will be 
determined later. Other matrices $h$ and $h'$ are defined in 
eqs.(\ref{eq:bpure}) and (\ref{eq:red}) with $p=1$,$q=0$,$\beta=1$ 
and $\delta=0$.

For example, let us consider the simplest ASEP model to make the notation 
clear. 
In this case, we can set $X\in V_{2}$ ({\it i.e.}$P_{1}=1,P_{2}=0$) and 
\[
H^{\rho\rho'}_{\sigma\sigma'}=h^{\rho\rho'}_{\sigma\sigma'}+
(h_{1})^{\rho}_{\sigma}
\]
It is obvious that we can reformulate the simplest ASEP  
model with $X\in V_{2^{l}}$. Anyway, that model can be considered as the 
simplest example in the following formalism. 
We want to answer the following question. 
\begin{quotation}
{\it Problem:~ What is the condition for the recursion relation}
\end{quotation}
\begin{equation}
f_{n}(X,\rho_{0},\rho_{1},\ldots,\rho_{n-1},1)=
f_{n-1}(X,\rho_{0},\rho_{1},\ldots,\rho_{n-1}) \quad for~n=1,2,\ldots \quad ? 
\label{basic}
\end{equation}
This relation comes from the recursion relation (\ref{eq:rrec}). 
We look for the answer to the above question. 
To begin with, we write down the master equation (\ref{scho})
explicitly. 
For $n=0$, we can easily obtain 
\BEA
\sum_{Y\in V_{s},\rho=0,1}H^{Y\rho}_{X1}f_{0}(Y,\rho)+f_{0}(X,1)&=&0, \\
\sum_{Y\in V_{s},\rho=0,1}H^{Y\rho}_{X0}f_{0}(Y,\rho)-f_{0}(X,1)&=&0,
\EEA
which corresponds to the initial condition for the distribution function 
in the following consideration. 
Next we consider the equation for $n=1$; 
\[
\sum_{Y,\sigma_{1}}
H^{Y\sigma_{1}}_{X\rho_{1}}f(Y,\sigma_{1},\rho_{2})+
h^{\sigma_{1}\sigma_{2}}_{\rho_{1}\rho_{2}}f(X,\sigma_{1},\sigma_{2})+
(h')^{\sigma_{2}}_{\rho_{2}}f(X,\rho_{1},\sigma_{2})=0.
\]
The independent elements of the above equation are the following 
three, 
\BEA
\sum_{Y,\sigma}H^{Y\sigma}_{X1}f_{1}(Y,\sigma,0)+f_{0}(X,0)&=&0, \\
\sum_{Y,\sigma}H^{Y\sigma}_{X0}f_{1}(Y,\sigma,0)-f_{0}(X,0)&=&0, \\
f_{1}(X,1,0)=f_{0}(X,1)+f_{0}(X,0), && \label{rec10}
\EEA
where we made use of the relation (\ref{basic}) and eq.(\ref{rec10}) takes 
the same form as that for the simplest ASEP model.
Similarly, for general $n(>0)$, we obtain the equation
\BEA
\sum_{Y,\sigma}H^{Y\sigma}_{X1}f_{n}(Y,\sigma,0,0,\ldots,0)
+f_{n-1}(X,0,\ldots,0)&=&0, \\
\sum_{Y,\sigma}H^{Y\sigma}_{X0}f_{n}(Y,\sigma,0,0,\ldots,0)
-f_{n-1}(X,0,\ldots,0)&=&0
\EEA
and 
\BEA\hspace{-1.0cm}
&&f_{n}(X,\sigma_{0},\ldots,\sigma_{i-1},1,0,\sigma_{i+2},\ldots,\sigma_{n})
\NN\\ \hspace{-1.0cm}&&\quad
=f_{n-1}(X,\sigma_{0},\ldots,\sigma_{i-1},1,\sigma_{i+2},\ldots,\sigma_{n})
+f_{n-1}(X,\sigma_{0},\ldots,\sigma_{i-1},0,\sigma_{i+2},\ldots,\sigma_{n}),
\NN\\ &&
\EEA
for $i=0,\ldots,n-1$. 
Upon setting new variables
\BE
x_{n}(X)=f_{n}(X,0,0,\ldots,0),\quad y_{n}(X)=f_{n}(X,1,0,\ldots,0), 
\quad n=0,1,2,\ldots,
\EE
we obtain the following equation
\BEA
\sum_{Y}H^{Y0}_{X1}x_{n}(Y)+\sum_{Y}H^{Y1}_{X1}y_{n}(Y)+x_{n-1}(X)&=&0,
\quad n=1,2,\ldots, \label{rec1}\\
\sum_{Y}H^{Y0}_{X0}x_{n}(Y)+\sum_{Y}H^{Y1}_{X0}y_{n}(Y)-x_{n-1}(X)&=&0,
\quad n=1,2,\ldots,\label{rec2}\\
\sum_{Y}H^{Y0}_{X1}x_{0}(Y)+\sum_{Y}H^{Y1}_{X1}y_{0}(Y)+y_{0}(X)&=&0 
\label{ini1}\\
\sum_{Y}H^{Y0}_{X0}x_{0}(Y)+\sum_{Y}H^{Y1}_{X0}y_{0}(Y)-y_{0}(X)&=&0 
\label{ini2}\\
y_{n}(X)=x_{n-1}(X)+y_{n-1}(X), && n=1,2,\ldots, \label{rec3}
\EEA
which is the condition for the recursion relation (\ref{basic}).

We consider the simplest ASEP model again as an  
example to make the notations 
clear. Now we put $X\in V_{1}$ instead of $V_{2}$ and can set 
$H^{Y\rho}_{X\sigma}=(h_{1})^{\rho}_{\sigma}$, 
where 
$(h_{1})^{0}_{0}=\alpha$, $(h_{1})^{0}_{1}=-\alpha$ 
$(h_{1})^{1}_{1}=\gamma$ and $(h_{1})^{1}_{0}=-\gamma$ at the 
left edge. 
We can ignore the argument $X$ in $x_{n}$ and $y_{n}$ in this 
notation. The equation (\ref{rec1})-(\ref{ini2}) reads simply 
\BEAN
\left(\begin{array}{cc}\alpha&-\gamma \\ -\alpha&\gamma\end{array}\right)
\left(\begin{array}{c}x_{n}\\ y_{n}\end{array}\right)&=&
\left(\begin{array}{c}x_{n-1}\\ -x_{n-1}\end{array}\right), \quad n=1,2,\ldots 
\\
\left(\begin{array}{cc}\alpha&-\gamma \\ -\alpha&\gamma\end{array}\right)
\left(\begin{array}{c}x_{0}\\ y_{0}\end{array}\right)&=&
\left(\begin{array}{c}y_{0}\\ -y_{0}\end{array}\right).
\EEAN
Therefore, we obtain
\[
x_{0}=\gamma+1,\quad,y_{0}=\alpha,\quad 
x_{n}={1\over\alpha}x_{n-1}+{\gamma\over\alpha}y_{n+1},
\]
which is compatible with the relation from (\ref{rec3}),
\[
y_{n}=x_{n-1}+y_{n-1},
\]
therefore, the wave functions in the simplest ASEP model satisfy the relation 
(\ref{basic}) as already known.

Next let us eliminate the distribution functions 
in eqs.(\ref{rec1})-(\ref{rec3}) and 
see the condition including only the parameters in the boundary matrix $H$. 
For this sake, we introduce an auxiliary variable $t$, functions 
\BEAN
\phi(P_{i};t)&=&\sum_{j=0}^{\infty}x_{j}(P_{i})t^{j}, \\
\psi(P_{i};t)&=&\sum_{j=0}^{\infty}y_{j}(P_{i})t^{j},
\EEAN
vectors 
\BEAN
{\vec v}_{n}&=&^{t}(x_{n}(P_{1}),y_{n}(P_{1}),\ldots,x_{n}(P_{s}),y_{n}(P_{s})), \\
{\vec \xi}(t)&=&^{t}(\phi(P_{1};t),\psi(P_{1};t),\ldots,
\phi(P_{s};t),\psi(P_{s};t)), \\
\EEAN
and $(2s)\times(2s)$ block diagonalized matrices
\BEAN
E_{1}=\left(\begin{array}{cccc}
e_{1}    &     &      &\bigzerou \\
         &e_{1}&      &          \\
         &     &\ddots&          \\
\bigzerol&     &      &e_{1}
\end{array}\right) &,&\qquad 
e_{1}=\left(\begin{array}{cc}
-1&0 \\ 1&0
\end{array}\right) \\
E_{2}=\left(\begin{array}{cccc}
e_{2}    &     &      &\bigzerou \\
         &e_{2}&      &          \\
         &     &\ddots&          \\
\bigzerol&     &      &e_{2}
\end{array}\right) &,&\qquad 
e_{1}=\left(\begin{array}{cc}
0&-1 \\ 0&1
\end{array}\right) .
\EEAN
We want to solve the recursion relation (\ref{rec1})-(\ref{rec3}).
Equation (\ref{rec3}) reads 
\BE
-t\phi(P_{i};t)+(1-t)\psi(P_{i};t)=y_{0}(P_{i}), \label{co1}
\EE
and eqs. (\ref{rec1})-(\ref{ini2}) mean
\BEA
H{\vec v}_{n+1}&=&-E_{1}{\vec v}_{n},              \label{co2}\\
H{\vec v}_{0}  &=&-E_{2}{\vec v}_{0}, \label{bc3} \label{co3}\\
H{\vec \xi}(t)&=&-tE_{1}{\vec \xi}(t)-E_{2}{\vec v}_{0}\label{co4}.
\EEA
The initial value $y_{0}(P_{i})$ can be determined from (\ref{bc3}) by the 
determinant of the $(2s-1)\times(2s-1)$ comatrix of $H+E_{2}$ like
\[
y_{0}(P_{i})=
\left|\begin{array}{cccccccc}
H^{P_{1}0}  _{P_{1}0} & H^{P_{1}1}_{P_{1}0}-1 & \cdots & H^{P_{i}0}_{P_{1}0} &
H^{P_{i+1}0}_{P_{1}0} & \cdots & H^{P_{s}0}_{P_{1}0}   & H^{P_{s}1}_{P_{1}0} \\
H^{P_{1}0}  _{P_{1}1} & H^{P_{1}1}_{P_{1}1}+1 & \cdots & H^{P_{i}0}_{P_{1}1} &
H^{P_{i+1}0}_{P_{1}1} & \cdots & H^{P_{s}0}_{P_{1}1}   & H^{P_{s}1}_{P_{1}1} \\
\dotmatrix{8}
H^{P_{1}0}  _{P_{i}0} & H^{P_{1}1}_{P_{i}0} & \cdots & H^{P_{i}0}_{P_{i}0} &
H^{P_{i+1}0}_{P_{i}0} & \cdots & H^{P_{s}0}_{P_{i}0}   & H^{P_{s}1}_{P_{i}0} \\
H^{P_{1}0}  _{P_{i+1}0}&H^{P_{1}1}_{P_{i+1}0}&\cdots& H^{P_{i}0}_{P_{i+1}0} &
H^{P_{i+1}0}_{P_{i+1}0}&\cdots&H^{P_{s}0}_{P_{i+1}0}& H^{P_{s}1}_{P_{i+1}0} \\
\dotmatrix{8}
H^{P_{1}0}  _{P_{s}0} & H^{P_{1}1}_{P_{s}0} & \cdots & H^{P_{i}0}_{P_{s}0} &
H^{P_{i+1}0}_{P_{s}0} & \cdots & H^{P_{s}0}_{P_{s}0}&H^{P_{s}1}_{P_{s}0}-1 \\
H^{P_{1}0}  _{P_{s}1} & H^{P_{1}1}_{P_{s}1} & \cdots & H^{P_{i}0}_{P_{s}1} &
H^{P_{i+1}0}_{P_{s}1} & \cdots & H^{P_{s}0}_{P_{s}1}& H^{P_{s}1}_{P_{s}1}+1
\end{array}\right|
. \]
Therefore, the condition for the recursion relation is the same as 
the constraint for the boundary matrix that there must be a solution 
$\vec{\xi}(t)$ 
for the simultaneous equations (\ref{co1})-(\ref{co4}). 
Let us introduce a $(2s+1)\times(2s+1)$ matrix 
$M_{ij}(t)$ for $i$ and $j$ ($i,j=1,2,\ldots,s$),
\BE
M_{ij}(t)=\left(\begin{array}{c|c} 
(H+tE_{1})' & {
\begin{array}{c}y_{0}(P_{1})\\ -y_{0}(P_{1})\\ \dots 
\\ y_{0}(P_{s}) \end{array}}\\
\hline
^{t}(-t{\vec e}_{P_{i}0}+(1-t){\vec e}_{P_{i}1})&y_{0}(P_{i}) \\
^{t}(-t{\vec e}_{P_{j}0}+(1-t){\vec e}_{P_{j}1})&y_{0}(P_{j})
\end{array}\right) ,
\EE
where $(2s-1)\times(2s)$ matrix $(H+tE_{1})'$ is obtained by eliminating  
the $(P_{s}1)$-th column vector from $(H+tE_{1})$ like 
\[
(H+tE_{1})'=\left(\begin{array}{ccccc} 
H^{P_{1}0}_{P_{1}0}-t&H^{P_{1}1}_{P_{1}0}&\ldots&
H^{P_{s}0}_{P_{1}0}  &H^{P_{s}1}_{P_{1}0} \\
H^{P_{1}0}_{P_{1}1}+t&H^{P_{1}1}_{P_{1}1}&\ldots&
H^{P_{s}0}_{P_{1}1}  &H^{P_{s}1}_{P_{1}1} \\
\dotmatrix{5}
H^{P_{1}0}_{P_{s}0}  &H^{P_{1}1}_{P_{s}0}&\ldots&
H^{P_{s}0}_{P_{s}0}-t&H^{P_{s}1}_{P_{s}0}
\end{array}\right),
\]
and ${\vec e}_{P_{i}j}$ is the unit vector in the $(P_{i}j)$-direction, 
for example,
\[
{\vec e}_{P_{1}0}=^{t}(1,0,\ldots,0),\quad 
{\vec e}_{P_{1}1}=^{t}(0,1,\ldots,0).
\]
Needless to say, there exist the other equivalent forms of $M_{ij}$, which 
are obtained by eliminating $(P_{i}1)$-th ($1\leq i\leq s$) column vector 
from $(H+tE_{1})$. 
With the above matrices $M_{ij}(t)$, the constraint 
can be written in the determinant forms as 
\BE
m_{ij}(t)=\det M_{ij}(t)=0 \label{eq:bdet}
\EE
for all $i$ and $j$ ($i,j=1,2,\ldots,s$) and arbitrary $t$.
The above equations are the main results in this section. 
\section{Calculation of Partition Function}\label{s:pf}
To calculate physical quantities such as the density currents and 
concentration, we have to obtain the form of the {\it partition function} 
$Z_{n}$:
\begin{equation}
Z_{n}=\sum_{X=P_{1},\ldots,P_{s}}\sum_{\{ \sigma_{i} =0,1 \} }
f_{n}(X,\sigma_{0},\sigma_{1},\ldots,\sigma_{n}).
\end{equation}
Several quantities can be extracted from this function. 
We calculate the partition function $Z_{n}$ with 
general $n$ exactly 
by means of a similar 
technique to that used in DDM's and SD's papers\cite{DDM,SD}. 
First, let us assume that we have already solved the master 
equation for some small $n$'s. It is nothing but a diagonalization 
problem for small dimensional matrices. 
Moreover, let us accept the following assumption. 
\begin{quotation}
{\it 
Assumption:~
With proper constants $s_{i}$ and $a_{i}$ ($i=1,2,\ldots,l$), 
the following recursion relation is satisfied.}
\BE
\sum_{X=P_{1},\ldots,P_{s}}
f_{n}(X,0,\ldots,0)-
\sum_{X=P_{1},\ldots,P_{s}}
f_{n+1}(X,0,\ldots,0,0)
=\sum_{i=1}^{l} s_{i}a_{i}^{n}\label{defcomp}.
\EE
\end{quotation}
Needless to say, the above relation should be checked in the following 
context. Although we have not succeeded in a general proof, 
we can show the above relation for specific models 
inductively. In the general treatment proceeded in this section, 
we provide that this relation is true. 
We prepare constants $u$ and $v$ defined by
\BEAN
u&=&\sum_{X=P_{1},\ldots,P_{s}}f_{0}(X,0)-\sum_{i}{s_{i}\over a_{i}}, \\
v&=&\sum_{X=P_{1},\ldots,P_{s}}f_{0}(X,1) \label{defgamma}.
\EEAN

Let us introduce the following value $Y_{n}(k)$, a similar quantity 
to HLP's in \S\ref{s:boundary}, defined by
\BEAN
&&Y_{n}(k)=\sum_{X\in V_{s}}\sum_{\{ \tau_{i}=0,1\} }
(1-\tau_{n})(1-\tau_{n-1})\cdots (1-\tau_{k})
f_{n}(X,\tau_{0},\tau_{1},\ldots,\tau_{n}),\\ 
&&\hspace{9.0cm} n\geq 0,0\leq k\leq n, \\
&&Y_{n}(n+1)=\sum_{X\in V_{s}}\sum_{\{ \tau_{i}=0,1\} }
f_{n}(X,\tau_{0},\tau_{1},\ldots,\tau_{n}),\quad n\geq 0, \\
&&Y_{-1}(0)=u,
\EEAN
and auxiliary functions $L_{p}(\lambda)$ as
\[
L_{p}(\lambda)=\sum_{n=p}^{\infty}\lambda^{n+1}
Y_{n}(n-p), \quad p=-1,0,1,\ldots
\]
with an extra parameter $\lambda$. By definition, we find
\[
L_{-1}(\lambda)=\sum_{n=-1}^{\infty}\lambda^{n+1}Y_{n}(n+1)
=u+\sum_{n=0}^{\infty}\lambda^{n} Z_{n}.
\]
It is not difficult to prove the following relations
\BEA
(1-\lambda)L_{-1}(\lambda)-L_{0}(\lambda)&=&u+v\lambda, 
\label{recl0} \\
L_{p+1}(\lambda)-L_{p}(\lambda)+\lambda L_{p-1}(\lambda)&=&
\lambda^{p+1}(Y_{p-1}(0)-Y_{p}(0)) \NN \\
&=&-\sum_{i}s_{i}a_{i}^{p-1},\quad{\rm for}\quad p\geq 0.
\label{recl}
\EEA
Now, we solve the recursion 
relation (\ref{recl}) under the initial condition (\ref{recl0}) and the 
condition that the first term in $L_{p}(\lambda)$ is proportional 
$\lambda^{p+1}$. 

For this sake, introducing a generating function 
$w(x,\lambda)$ with another extra variable $x$ as 
\[
w(x,\lambda)=\sum_{p=0}^{\infty}L_{p-1}x^{p},
\]
the recursion relation (\ref{recl}) becomes 
\[
(1-x+\lambda x^{2})
w(x,\lambda)=L_{-1}+x(L_{0}-L_{-1})+\sum_{p=1}^{\infty}x^{p+1}\lambda^{p}
\left(\sum_{i}s_{i}a_{i}^{p-1}\right)
\]
giving the solution 
\BEA
L_{n}&=&
\frac{\omega_{1}^{n+2}-\omega_{2}^{n+2}}{\omega_{1}-\omega_{2}}L_{-1}+
\frac{\omega_{1}^{n+1}-\omega_{2}^{n+1}}{\omega_{1}-\omega_{2}}(L_{0}-L_{-1}) 
\NN \\
&&+\sum_{i}s_{i}\lambda
\left[
\frac{\omega_{1}^{n+1}}{(\omega_{1}-\omega_{2})(\omega_{1}-a_{i}\lambda)}+
\frac{\omega_{2}^{n+1}}{(\omega_{2}-\omega_{1})(\omega_{2}-a_{i}\lambda)}+
\frac{(a_{i}\lambda)^{n+1}}
{(a_{i}\lambda-\omega_{1})(a_{i}\lambda-\omega_{2})}
\right]  \NN \\
&&
\EEA
with functions
\[
\omega_{1}=\frac{1-\sqrt{1-4\lambda}}{2},\quad 
\omega_{2}=\frac{1+\sqrt{1-4\lambda}}{2}.
\]
Demanding the general condition 
$L_{p}(\lambda)\sim\lambda^{p+1}$, we obtain a simple equation 
\[
-\omega_{2}L_{-1}-(L_{0}-L_{-1})-\sum_{i=1}^{l}s_{i}\lambda
{1\over\omega_{2}-a_{i}\lambda}=0
\]
and the initial condition (\ref{recl0}). Finally, we see
\BEAN
L_{-1}(\lambda)&=&
\left({\omega_{1}\over\lambda}\right)^{2}
\left[ u+v\lambda-\sum_{i=1}^{l}s_{i}\lambda
{1\over\omega_{2}-a_{i}\lambda}\right] \NN \\
&=&u\left({\omega_{1}\over\lambda}\right)^{2}+
v\lambda\left({\omega_{1}\over\lambda}\right)^{2}-
\lambda\left({\omega_{1}\over\lambda}\right)^{3}\sum_{i}s_{i}\\ &&
\quad -\lambda^{2}\left({\omega_{1}\over\lambda}\right)^{4}\sum_{i}s_{i}a_{i}-
\lambda^{3}\left({\omega_{1}\over\lambda}\right)^{5}\sum_{i}s_{i}a_{i}^{2}
-\cdots.
\EEAN

Expanding 
$L_{-1}(\lambda)$ as for $\lambda$ with helpful relations 
\[
\left({\omega_{1}\over\lambda}\right)^{m}=\sum_{n=0}^{\infty}
m\frac{(2n+m-1)!}{n!\,(n+m)!}\lambda^{n},\quad m=1,2,\ldots
\]
gives the expressions
\BEA
&&Z_{n}=\left[u+{v\over 2}{n+1\over 2n+1}\right] Z_{n}^{(0)}
+Z_{n}^{(1)}, \NN \\
&&Z_{n}^{(0)}={(2n+2)!\over (n+1)!\,(n+2)!},\quad 
Z_{n}^{(1)}=
-\sum_{i=1}^{l}{s_{i}\over a_{i}}R_{n}(a_{i}), \label{partition}
\EEA
with a function $R_{n}(a)$ defined by
\BE
R_{n}(a)=\sum_{m=1}^{n}(m+2){(2n-m+1)!\over (n+2)!\,(n-m)!}a^{m}.
\label{defr}
\EE
The above is the exact form of the partition functions. 

Let us investigate the asymptotic behavior of $R_{a}(x)$ defined in 
(\ref{defr}) with large $n$ for later convenience. 
We define functions
\[
r_{n,m}(a)=(m+2)\frac{(2m-m+1)!}{(n+2)!\,(n-m)!}a^{m},\quad m=1,2,\ldots,n.
\]
\begin{itemize}
\item Region $a<2$ \\
It is not difficult to see that the asymptotic behavior of the function 
$r_{n,m}(a)$ for large $n$ is
\BE
r_{n,1}(a)={3\over 2}aZ_{n}^{(0)},\quad 
r_{n,m}(a)\sim m\left({a\over 2}\right)^{m}r_{n,1}(a).
\EE
Therefore, the asymptotic form of the function $R_{n}(a)$ for large $n$ can
be written as
\BE
R_{n}(a)\sim c(a)\times Z_{n}^{(0)},\label{eq:lt2}
\EE
where the function $c(a)$ is independent of $n$
\item Region $a>2$ \\
For large $n$, because 
$R_{n}(a)$ is dominated by $max_{1\leq k\leq n}~r_{n,k}(a)$, we obtain 
\BEA
R_{n}(a)&\sim&{1\over n^{3/2}}\tau(a)^{n}, \label{eq:gt2}\\
\tau(a)&=&a^{a\over a-1}
\left({a-2\over a-1}\right)^{-{a-2\over a-1}}
\left(1+{a-2\over a-1}\right)
^{1+{a-2\over a-1}}.
\EEA
\end{itemize}
We will make use of the above expression in the next section.
\section{Density Current and Concentration}\label{s:phase}
In this section, we calculate some physical quantities and 
draw some {\it phase} {\it diagrams}, which  
just means the diagrams of the difference of the asymptotic forms 
of the density current and one-point function. 

First, we consider the density current 
\[
J=\lim_{n\rightarrow\infty}\sum_{X}\sum_{j\neq k}\sum_{\{ \sigma \}}
{
f_{n}(X,\sigma_{0},\cdots,\sigma_{k-1},1,0,\sigma_{k+2},\cdots,\sigma_{n})
\over Z_{n}},
\]
which is actually independent of $k(=0,1,\cdots,n)$ and corresponds to
\BE
J=\lim_{n\rightarrow\infty}{Z_{n-1}\over Z_{n}}. \label{eq:current}
\EE
From the asymptotic form of the function $R_{n}(a)$, 
(\ref{eq:lt2}) and (\ref{eq:gt2}), we can calculate the current as 
follows. It must be remarked that the current does not depend on 
$s_{i}$ but only on the eigenvalues $a_{i}$.
\begin{itemize}
\item Case 1:~$a_{i}>2$ for some $i$'s. \\
Let us assume
\[
a_{1}>a_{2}>\cdots >a_{k}>2>a_{k+1}>\cdots >a_{l}.
\]
From the asymptotic form of $R_{n}(a)$, it is not so difficult to show
\[
1/J=a_{1}^{a_{1}\over a_{1}-1}
\left({a_{1}-2\over a_{1}-1}\right)^{-{a_{1}-2\over a_{1}-1}}
\left(1+{a_{1}-2\over a_{1}-1}\right)
^{1+{a_{1}-2\over a_{1}-1}}.
\]
We must remark $J<1/4$ in this case.
\item Case 2:~$a_{i}<2$ for all $i(=1,2,\ldots,l)$. \\
According to eq.(\ref{eq:lt2}), the partition function 
$Z_{n}$ is proportional to $Z_{n}^{(0)}$ in this case. Therefore, 
we can show
\[
J=\lim_{n\rightarrow\infty}{Z^{(0)}_{n-1}\over Z^{(0)}_{n}}={1\over 4}.
\]
Hence, the density current in this region is not so sensitive to 
the parameters as that in the previous region. 
We call this region {\it maximal} {\it current} {\it phase}\cite{DDM,SD}.
\end{itemize}

\begin{figure}
  \epsfxsize = 6 cm   
  \centerline{\epsfbox{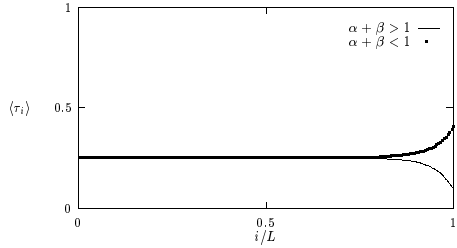}}
  \caption{Concentration ($\expect{\tau_{i}}$) in the low density phase. 
$\alpha$ and $\beta$ correspond to those in the simplest ASEP model 
with $p=1$,$\gamma=\delta=0$. This notation is also used in Figs. 
\ref{f:high} and \ref{f:max}.}
  \label{f:low}
\end{figure}

\begin{figure}
  \epsfxsize = 6 cm   
  \centerline{\epsfbox{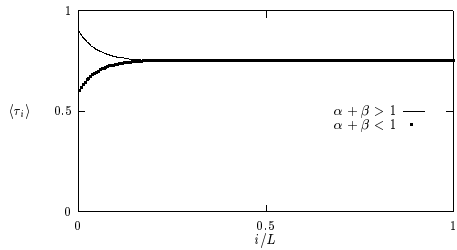}}
  \caption{Concentration in the high density phase.}
  \label{f:high}
\end{figure}

\begin{figure}
  \epsfxsize = 6 cm   
  \centerline{\epsfbox{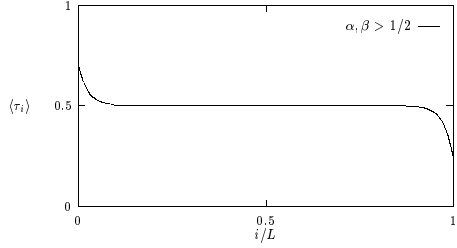}}
  \caption{Concentration in the maximal current phase. }
  \label{f:max}
\end{figure}

We can say nothing more from only the density current. However,  
the above former region is classified further into several phases. 
Let us consider the cases with two eigenvalues, \IE $l=2$. In the 
cases with two eigenvalues, the concentration, \IE one-point function, 
\BEA
\expect{\sigma_{n-m}}_{n}&\teigi&{1\over Z_{n}}
\sum_{X=P_{1},\ldots,P_{s}}\sum_{\{ \sigma_{i} =0,1 \} }
\delta_{\sigma_{n-m},1}f_{n}(X,\sigma_{0},\sigma_{1},\ldots,\sigma_{n})
\NN \\
&=&{1\over Z_{n}}
\sum_{j=0}^{m}{(2j)!\over j!\,(j+1)!}Z_{n-j-1}.
\EEA
separates the 
{\it low} and 
{\it high} {\it density} phases\cite{DDM,SD}. 
The shapes of the concentrations in the 
low and high density and maximal current phases are shown 
in Figs.\ref{f:low},\ref{f:high} and \ref{f:max}, respectively. 
As shown in a general scheme done in \cite{SD,KKSS} without the 
mean-field approximation, 
systems with just two eigenvalues can undergo no other {\it phase  
transition} other than those in Figs.\ref{f:low}-\ref{f:max}. 
The problems on systems with more than three eigenvalues should be 
open. 
%
%
\section{$V_{2}$-Model}\label{s:v_2}
We consider the special model with $P_{i}=0,1$ in some detail. 
The boundary matrix is given by 12 independent parameters 
$H^{ij}_{kl}$ with $\sum_{k,l}H^{ij}_{kl}=0$. The initial condition 
$H{\vec v}_{0}=-E_{2}{\vec v}_{0}$ is explicitly solved as 
\BEA
y_{0}(0)&=&
\left|\begin{array}{ccc}
H^{00}_{00}&H^{10}_{00}&H^{11}_{00}  \\
H^{00}_{10}&H^{10}_{10}&H^{11}_{10}-1\\
H^{00}_{11}&H^{10}_{11}&H^{11}_{11}+1
\end{array}\right|, \\
y_{0}(1)&=&
\left|\begin{array}{ccc}
H^{00}_{00}&H^{01}_{00}-1&H^{10}_{00}\\
H^{00}_{01}&H^{01}_{01}+1&H^{10}_{01}\\
H^{00}_{10}&H^{01}_{10}  &H^{10}_{10}
\end{array}\right| .
\EEA
Upon setting the function $m(t)$ as
\BE
m(t)=\left|\begin{array}{ccccc}
H^{00}_{00}-t&H^{01}_{00}&H^{10}_{00}  &H^{11}_{00}& y_{0}(0)\\
H^{00}_{01}+t&H^{01}_{01}&H^{10}_{01}  &H^{11}_{01}&-y_{0}(0)\\
H^{00}_{10}  &H^{01}_{10}&H^{10}_{10}-t&H^{11}_{10}& y_{0}(1)\\
-t           &1-t        &0            &0          & y_{0}(0)\\
0            &0          &-t           &1-t        & y_{0}(1)
\end{array}\right| , \label{v2}
\EE
the condition is shown $m(t)=0$ independent of $t$, which gives 
3 relations among the 12 parameters $H^{ij}_{kl}$. 
%
As an example, we concentrate the model with
\[
H^{01}_{11}=a,\quad H^{00}_{10}=b,\quad H^{10}_{01}=p, 
\quad H^{01}_{10}=q, \quad H^{00}_{11}=r,
\]
and the other parameters $H^{ij}_{kl}=0$ for $(i,j)\neq(k,l)$. 
The equation gives the relation
\begin{equation}
a+q=ap, \quad (r+b)(ap-r-b)+pr=0.
\end{equation}
For instance, the model with $a=b=1$,$p=2$,$q=1$ and $r=1+\sqrt{2}$ 
satisfies this condition.
It is obvious that the simplest ASEP model ($a=b$,$p=1$ and $q=r=0$)
is the simplest example. 

Let us consider the eigenvalue problem. 
In this case, we see that there are two eigenvalues ($l=2$) as well as the 
simplest ASEP model and the eigenvalues $a_{1}$ and $a_{2}$ are 
calculated as
\[
a_{1}={1\over b+r},\quad a_{2}=1.
\]
Therefore, we see that if the parameters satisfy $1/(b+r)\leq 2(>2)$, 
the system indicates the {\it maximal current phase} ({\it low density phase}).
%
%
\section{Summary and Discussion}\label{s:conclusion}
As we have seen, possible integrable boundary interaction should satisfy  
eq. (\ref{eq:bdet}). The density current and concentration can be 
expressed in terms of the partition functions exactly determined by 
the recursion relations (\ref{eq:lrec})-(\ref{eq:rrec}). 
The extension for general $\beta$ and $\gamma$ at the right edge and 
$p$ and $q$ in the bulk is straightforward. Because  
stochastic systems are equivalent to ferromagnetic spin chains,  
their behavior highly depends on the boundary conditions. 

We must remark several points. First, 
the fact that an eigenvalue $a_{2}$ equals to $1$ 
originates in the choice that we take an extraction rate $1$ 
at the right edge.
Secondly, some of the readers may wonder;
\begin{quotation}
{\it If we hide or ignore the first site, 
does the system reduce to the simplest  
ASEP model? }
\end{quotation}
In fact, the answer is {\it no}. There are two reasons or evidences. 
First, we must pay 
attention that $a_{1}$ is not equal to just the sum of the probability 
$\sum_{ij}H^{i0}_{j1}$ because of the conservation of probability at the 
left edge. Secondly, even if we focus on the 
simplest ASEP model, 
we cannot obtain the precise results by hiding the left edge site
\cite{DDM,SD}. 
This fact comes from the recursion relations (\ref{eq:lrec})-
(\ref{eq:rrec}). 
The form of $a_{1}$ indicates an effect of bulk correlation, 
though it is not so clear because we have set $p=1$
in this case. 

In general, the current $J$ of the system does not depend on 
the parameters $H^{Y\rho}_{X\sigma}$ explicitly but on 
the eigenvalues $a_{i}$. It is still unknown whether there exists 
an interesting model with $X\in V_{s}~(s\geq 3)$, especially 
$X\in V_{2}\otimes V_{2}\otimes\cdots\otimes V_{2}$. 
%
%
%
\begin{center}
{\bf Acknowledgements}
\end{center}
{\small  
The author would like to thank the members of the theoretical 
physics group in Univ. of Cologne and Univ. of Bonn for the 
discussions, comments and their hospitality. 
He acknowledges R.~Rauph 
for reading the manuscript. The useful comments on the previous 
version sent by G.~Sch{\"u}tz are much appreciated. 
This work is financially 
support by DFG, SFB341(KL645/3-1) and Soryushisyogakukai.}

\end{document}